\documentclass{article}

\usepackage{PRIMEarxiv}
\usepackage{amsmath}
\usepackage[utf8]{inputenc} 
\usepackage[T1]{fontenc}    
\usepackage{hyperref}       
\usepackage{url}            
\usepackage{booktabs}       
\usepackage{amsfonts}       
\usepackage{amssymb}
\usepackage{pifont}
\usepackage{nicefrac}       
\usepackage{microtype}      
\usepackage{lipsum}
\usepackage{amsmath, nccmath}
\usepackage{xcolor}
\definecolor{mintbg}{rgb}{.63,.79,.95}
\definecolor{red}{rgb}{1,0,.0}
\definecolor{yellow}{rgb}{1,1,0}
\colorlet{lightmintbg}{mintbg!30}
\colorlet{lightred}{red!10}
\colorlet{lightyellow}{yellow!40}

\usepackage{geometry}
\usepackage{fancyhdr}       
\usepackage{graphicx}       
\graphicspath{{media/}}     
\title{Transport-Embedded Neural Architecture: Redefining the Landscape of physics aware neural models in fluid mechanics}

\author{
  Amirmahdi Jafari \\
  Research assistant \\
  Sharif University of Technology \\
  \texttt{jafary.am@gmail.com} \\
   \\
}

\begin{document}
\maketitle

\begin{abstract}
This work introduces a new neural model which follows the transport equation by design. A physical problem, the Taylor-Green vortex, defined on a bi-periodic domain, is used as a benchmark to evaluate the performance of both the standard physics-informed neural network and our model (transport-embedded neural network). Results exhibit that while the standard physics-informed neural network fails to predict the solution accurately and merely returns the initial condition for the entire time span, our model successfully captures the temporal changes in the physics, particularly for high Reynolds numbers of the flow. Additionally, the ability of our model to prevent false minima can pave the way for addressing multiphysics problems, which are more prone to false minima, and help them accurately predict complex physics.
\end{abstract}


\section{Introduction}
Transport processes are integral to a variety of flow problems, describing the movement of quantities like mass or energy within the flow. These equations are robust local representations of fundamental conservation laws in physics and appear in numerous applications, from heat transfer\cite{bergman2011introduction} to drug delivery in biofluidic flows\cite{longest2019use}. Depending on the specific quantity being transported, they are known by different names, such as convection-diffusion equations for species and energy transport or Navier-Stokes equations for momentum transport. With recent advancements in computational fluid dynamics (CFD), several methods, including finite element\cite{lewis2004fundamentals}, finite volume\cite{moukalled2016finite} and meshless techniques\cite{katz2009meshless} have been developed. However, these methods struggle to effectively integrate real-world data into their frameworks. This limitation is particularly problematic for scenarios where existing theories do not fully capture flow behavior such as high-Reynolds number flows\cite{eivazi2022physics} or where data collection is difficult, costly, or noisy\cite{jin2021nsfnets}. Additionally, they cannot solve inverse problems with relatively low computational costs. To overcome these challenges and improve accuracy in data-sparse conditions, physics-aware models have been proposed. These models leverage mathematical physics and data-driven techniques from machine learning to address more realistic physics problems compared to traditional numerical methods.
These models are categorized into three groups: the first group includes models that are guided by physics without directly embedding it into their structure. They rely on data-dependent supervised optimization, which limits their ability to generalize or extrapolate their understanding of physics beyond the spatiotemporal data on which they were trained\cite{huang2021machine,lee1993fluid,faroughi2022meta}. Also the majority of the real-world problems aren't big data-oriented and extracted datasets from experiments cannot support all possible conditions.
The second group, known as physics-informed neural models, uses a loss function comprising residuals of the partial differential equations they aim to solve, along with boundary and initial conditions provided as supervised input data. These models compute the derivatives of the outputs with respect to their inputs using automatic differentiation, eliminating the need for discretization or mesh generation\cite{jin2021nsfnets,eivazi2022physics,wang2017physics}.
Both of the above groups are limited by issues related to stability, convergence, and the inability to perform successful optimization under certain constraints.
The last group consists of models that directly embed physics formulations into their architecture. The capability to encode complex equations with non-linearities simplifies the training process, enabling these models to perform well in sparse data situations without concerns about generalizing physics beyond the spatiotemporal boundaries\cite{chen2018neural,wang1994deterministic,cranmer2020lagrangian,trask2022enforcing}. 
Currently, Physics-aware neural models are neither as accurate nor as fast as traditional computational fluid dynamics (CFD) techniques, however they offer better scalability and flexibility. This is primarily due to the non-convex, high-dimensional loss functions that arise in fluid dynamics applications. In this experiment, a standard benchmark for evaluating the accuracy and stability of numerical methods in solving the Navier-Stokes equations is used to assess the performance of standard physics-informed neural networks, as introduced by \cite{raissi2019physics} (referred to as vanilla PINN), and our physics-embedded neural architecture. We compare their accuracy across a range of flow scenarios.
\section{Method}
\subsection{Problem Overview}
\textbf{Transport Equation}\hspace{0.2cm}The fundamental laws of mechanics, such as the second Newton's law which have been initially developed for particle systems can be modified to describe the same laws for continuum media through Reynold's transport theorem\cite{reynolds1900papers}. By this theorem, we can arrive at the scalar transport(or convection-diffusion) equations. Here we are encountered by a physical constraint on the model represented by a partial differential equation.
\begin{equation}\label{eq:transportEq}
\partial_t \phi + \partial_k (\phi u_k) = \partial_k(\Gamma \partial_k \phi ), \hspace{0.2cm} \phi \in \mathbb{R} , \bold{u} \in \mathbb{R}^{n}, \hspace{0.2cm} k = 1,..,n \end{equation}
Here, $n$ refers to the spatial dimensions, and index $k$ follows the Einstein summation convention, while $\Gamma$ denotes the diffusion constant. The first two terms on the left-hand side (LHS) represent the rate of change within the system combined with the advection of quantities, which move along with the system (at velocity $\bold{u}$). This should be equal to the diffusing quantity, considered stationary relative to the system, as described by Fick's law \cite{paul2014fick}. Such equations describe physical phenomena where quantities like energy or species are transported within a system.

\textbf{Incompressible Navier-Stokes Equations}\hspace{0.17cm} A famous subset of transport equations, are Navier-Stokes equations which demonstrate the momentum conservation in a viscous system.

\begin{equation}\label{NS}
\left\{\begin{matrix}
\partial_t u_i + \partial_j (u_i u_j) = -\partial_i p + \frac{1}{Re}\partial_i \partial_j u_i, \hspace{0.2cm} \bold{u} \in \mathbb{R}^{n}, \hspace{0.2cm} i,j = 1,..,n
 \\ \hspace{-7.45cm} \partial_i u_i = 0
 \end{matrix}\right.\end{equation}
where $\bold{u} \in H$ and $H$ is a space of divergence free vector fields. 
The solution of the equations is the flow's velocity and pressure fields at any moment in a time interval. It is usually studied in three spatial dimensions and one time dimension, although two (spatial) dimensional and steady-state cases are often used as models, and higher-dimensional analogues are studied in both pure and applied mathematics. Once the velocity field is calculated, other quantities of interest such as pressure or temperature may be found using dynamical equations and auxiliary relations. 
In the following section, we will introduce a problem which is used as benchmark to evaluate numerical models. 

\subsection{Transport on a $\mathbb{T}^2$}
As a classical problem in fluid mechanics, investigating turbulence and transport on a bi-periodic domain has been of interest since the Taylor-Green vortex flow was first introduced by Taylor et al.\cite{taylor1937mechanism} and later investigated by Brachet et al.\cite{brachet1983small}. The Taylor-Green vortex problem is a canonical flow which has become a challenging benchmark test for numerical methods, as the geometry is simple while the phenomena represented are complex. The flow represents an initially vortex that decays with the formation of a cascade of progressively smaller vortex structures due to vortex stretching mechanisms caused by viscosity. This flow features transient and fully turbulent behaviour, and the velocity field remains anisotropic over a period of time depending on the Reynolds number. Similar to Taylor-Green experiment, our experiment, is a flow governed by Navier-Stokes equations, bounded inside a 1 by 1 periodic domain, also known as a unit Toroid, given a smooth initial condition as follows:
\begin{equation}\label{VorttransportEq}
\textbf{u}_0 = \textbf{u}(x)_{t=0} = \begin{bmatrix}
 \text{cos}(2\pi x)\text{sin}(2\pi y)\\
 -\text{cos}(2\pi y)\text{sin}(2\pi x)
\end{bmatrix} 
\end{equation}

According to \cite{brzezniak2016existence}, no matter the initial conditions, as long as it is smooth, Navier-Stokes problem on a $\mathbb{T}^2$ will always have a bounded and unique solution.


\textbf{Vorticity Transport}\hspace{0.2cm}
The vorticity transport equation can be derived from the Navier-Stokes equation by applying the curl operator. Taking the curl of the Navier-Stokes equation yields the following form:

\begin{equation}\label{eq:VorttransportEq}
\frac{\partial \omega_i}{\partial t} + u_i \frac{\partial \omega_j}{\partial x_i} - \omega_i \frac{\partial u_j}{\partial x_i} + \omega_i \left( \frac{\partial u_j}{\partial x_j} \right) + u_i \left( \frac{\partial \omega_j}{\partial x_j} \right) = -\epsilon_{ijk} \partial_j \partial_k \left( \frac{p}{\rho} \right) + \omega_j \frac{\partial u_i}{\partial x_j} + \Gamma \frac{\partial^2 \omega_i}{\partial x_j \partial x_j}
\end{equation}

Here, \(\omega = \nabla \times \mathbf{u}\) represents the vorticity, where \(\mathbf{u}\) is the velocity field, and \(\Gamma = \frac{1}{Re}\) accounts for the diffusion effects, which are associated with the viscosity in the fluid. The vorticity equation describes the evolution of the vorticity vector \(\omega_i\) under the influence of fluid flow, viscous diffusion, and pressure gradients.
Given that the velocity field is incompressible, which implies \(\nabla \cdot \mathbf{u} = 0\), certain terms simplify. In the case of a planar velocity field, the divergence conditions \(\frac{\partial \omega_j}{\partial x_j} = 0\) and \(\frac{\partial u_j}{\partial x_j} = 0\) hold, which lead to the cancellation of terms like \(\omega_i \frac{\partial u_j}{\partial x_j} = 0\). Additionally, the curl of the pressure gradient term, \(\frac{p}{\rho}\), vanishes under the assumptions of either a uniform density field or a condition where the pressure and density gradients are orthogonal.
With these assumptions, the vorticity transport equation simplifies to:

\begin{equation}\label{eq:VorttransportEq2}
\frac{\partial \omega_i}{\partial t} + u_j \frac{\partial \omega_i}{\partial x_j} = \omega_j \frac{\partial u_i}{\partial x_j} + \Gamma \frac{\partial^2 \omega_i}{\partial x_j \partial x_j}
\end{equation}

For two-dimensional or planar flows, only the component of vorticity normal to the plane (i.e., perpendicular to the flow) remains non-zero. As a result, the term \(\omega_j \frac{\partial u_i}{\partial x_j} = 0\) vanishes due to the lack of variation in the velocity components in the perpendicular direction.
Taking all the above considerations into account, the vorticity transport equation reduces further to the scalar form:

\begin{equation}\label{eq:VorttransportEq3}
\frac{\partial \omega}{\partial t} + u_j \frac{\partial \omega}{\partial x_j} = \Gamma \frac{\partial^2 \omega}{\partial x_j \partial x_j}
\end{equation}

This form represents the scalar advection-diffusion equation for vorticity, where the left-hand side describes the advection of vorticity by the fluid motion, and the right-hand side corresponds to the viscous diffusion of vorticity. This simplified vorticity transport equation is particularly useful in analyzing two-dimensional incompressible flows, where the vorticity is confined to the plane and evolves due to both the flow dynamics and diffusion effects.
\subsection{Embedding Scalar Transport into Neural Approximations}
Let \( v \) represent the output vector predicted by a neural model for a given transport problem. We redefine the concept of divergence by incorporating the temporal component, denoted as \(\text{Div}\). For an \(N\)-dimensional vector \(v\), the divergence is expressed as \(\text{Div}(v) = \frac{\partial v_0}{\partial t} + \frac{\partial v_i}{\partial x_i}\). To embed convection, we construct the convection-embedding tensor \(A\) as follows:
\begin{equation}
A_{ij} = \epsilon_{ijk} v_{k}
\end{equation}
where \(\epsilon_{ijk}\) is the Levi-Civita tensor. The row-wise divergence of \(A\) yields:
\begin{equation}
T = \partial_i A_{ij} = \partial_i (\epsilon_{ijk} v_k) = \epsilon_{ijk} \partial_i v_k
\end{equation}

\textbf{Lemma 1: \(T\) is divergence-free}

\textbf{Proof}: Taking the divergence of \(T\), we obtain:
\begin{equation}
\text{Div}(T) = \epsilon_{ijk} \partial_j \partial_i v_k
\end{equation}
Interchanging the indices \(i\) and \(j\) leaves the divergence unchanged, but applying this interchange results in:
\begin{equation}
\text{Div}(T) = \epsilon_{jik} \partial_i \partial_j v_k = -\text{Div}(T)
\end{equation}
Using the symmetry property of continuous functions, \(\partial_i \partial_j = \partial_j \partial_i\), and the antisymmetric property of the Levi-Civita tensor, \(\epsilon_{jik} = -\epsilon_{ijk}\), we conclude that \(\text{Div}(T) = 0\).

Next, considering the first component of \(T\), denoted as \(T_0\), we define the diffusion-embedding matrix \(D\) as:
\begin{equation}
D_{ij} = \delta_{ij} T_0
\end{equation}
Taking the row-wise divergence of \(D\) and multiplying by a diffusion constant results in \(R = \Gamma \partial_i D_{ij}\). We now demonstrate that \(M = T + R\) represents the desired output for training the neural model.

\textbf{Lemma 2: \(M\) satisfies the transport equation}

\textbf{Proof}: We express the output \(M\) for training as \(M = R + T\). By defining the components of \(M\) as \(M_0 = \phi\) and \(M_i = \phi u_i\) for \(i = 1, \ldots, n\), and noting that the first component of \(R\) is zero, we have \(M_0 = T_0 = \phi\). Thus:
\begin{equation}
\text{Div}(M) = \text{Div}(R + T) = \text{Div}(T) + \text{Div}(R) = \frac{\partial \phi}{\partial t} + \frac{\partial (\phi u_i)}{\partial x_i} = 0 + \frac{\partial (\Gamma \frac{\partial \phi}{\partial x_i})}{\partial x_i}
\end{equation}
For matrix demonstration check appendix A.

\section{Related Works}
\textbf{Enforcing periodicity}\hspace{0.2cm} Periodic boundary conditions are a mathematical approach used in mechanics and fluid mechanics to model infinite or repeating systems by assuming that the boundaries of a simulation domain are connected. This means that when a particle, wave, or flow exits one side of the domain, it re-enters from the opposite side as if the system is continuous and unbounded.
In fluid mechanics, periodic boundary conditions are used to simulate flows in situations where the flow pattern is expected to repeat over a certain distance, such as in channels or across arrays of obstacles. This technique simplifies the computational effort by reducing the size of the computational domain while maintaining the overall characteristics of the flow.
There has been numerous works on how to work model periodic functions in deep neural networks, whether as a boundary condition or prediction of a periodic output function. Some papers add a loss function to the overall loss as to making the condition on periodic axis identical, assuming the $x_i$ axis to be the periodic axis, $\mathcal{L}_{PBC} := \left\|  \bold{\mathbf{y}_{x_{i}=0} - \mathbf{y}_{x_{i}=T}} \right \|_{\bold{\partial \Omega}}$ with $T$ being the periodic constant\cite{kumar2024physics, shah2024physics, krishnapriyan2021characterizing}. Another approach is to enforce boundary condition with prior knowledge to the neural model. In this method  Putting periodic trigonometric\cite{peng2020accelerating, kast2024positional,dong2021method} or polynomial\cite{dong2021method} feedforward in the middle between input and the rest of the neural model(cite) which guarantees periodicity has been discussed in literature. Also using periodic activation function has also been investigated as a way of helping neural models understand the nature of periodic functions. 
\begin{figure}[!ht]%
    \centering
    {{\includegraphics[width=15.5cm]{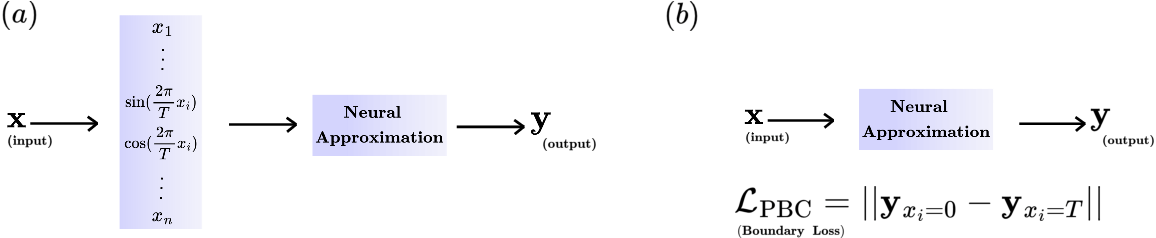}}}%
    \caption{Two popular approaches to deal with periodic boundary conditions, (a) Prior-knowledge and (b) direct incorporation of a loss into the final loss function}%
    \label{PD}%
\end{figure}

\textbf{Failures of Physics-informed neural models}\hspace{0.2cm}
The application of physics-informed neural networks (PINNs) to solve convection-diffusion problems has garnered significant interest since the concept was introduced by Raissi et al. in 2019 \cite{raissi2019physics}. While these models have achieved considerable success, certain challenges persist, particularly in specific training scenarios. These challenges often arise due to the highly non-convex nature of the loss functions in these models, leading to unbalanced gradients during training. Failures in training PINNs can generally be categorized into three types:
\begin{enumerate}
    \item \textbf{Trivial Solutions:} In some cases, the models converge to trivial solutions such as \( \bold{y}(x,t) = 0 \), which is technically a solution for convection-diffusion partial differential equations (PDEs), but not a useful one.
    
    \item \textbf{Static Solutions:} Other models converge to \( \bold{y}(x,t) = \bold{y}(x,0) \) or \( \frac{\partial \bold{y}}{\partial t} = 0 \), implying that the initial condition remains the solution throughout the time span, or the solution does not evolve over time, which is also uninformative.
    
    \item \textbf{Insufficient Accuracy:} A third category involves models that successfully capture the underlying physics and temporal progression of the problem but fail to predict the solution with sufficient accuracy.

\end{enumerate}

Krishnapriyan et al. \cite{krishnapriyan2021characterizing} explored the failure modes of convection, diffusion, and convection-diffusion problems under periodic boundary conditions. They found that in certain scenarios, such as those with low diffusion coefficients, PINNs struggled to capture the underlying physics and instead overfitted to trivial solution. They suggested approaches such as curriculum learning and sequence-to-sequence (seq2seq) learning to address these issues.
Ang et al. \cite{ang2023physics} reported difficulties in using PINNs to model fluid flow over a cylinder at low Reynolds numbers. By adjusting hyperparameters such as the number of layers and nodes per layer, they were able to improve the results, though the final accuracy was still far from the ground truth.

For incompressible flows, satisfying the divergence-free condition of the velocity field is critical. Wang et al. \cite{wang2021understanding} investigated gradient flow pathologies in lid-driven cavity flows. By introducing a stream function, which is the inverse curl of the velocity field, they automatically ensured incompressibility and achieved high accuracy.
Rao et al. \cite{rao2020physics} employed a streamfunction-pressure formulation and incorporated the stress tensor in place of viscous diffusion terms within their PDE formulation to predict flow over a cylinder at moderate Reynolds numbers . Similarly, Richter et al. \cite{richter2022neural} developed a neural network that respects the continuity equation without explicitly incorporating it into the loss function. Instead, they used network parameterization to avoid converging to static solutions. The continuity equation, a fundamental principle of mass conservation, is a critical component in fluid dynamics.

Another approach to overcoming these issues was presented by Leiteritz et al. \cite{leiteritz2021avoid}, who introduced penalty terms in the loss function to avoid static solutions. Basir et al. \cite{basir2022investigating} proposed Physics and Equality Constrained Artificial Neural Networks (PECANNs), which are based on constrained optimization methods. They introduced auxiliary variables to reduce the final error in the neural network, inspired by techniques from finite element methods, which effectively reduce the order of the PDE. In their method, rather than setting the residuals of each equation directly to zero, they constrained them within a small positive range, \( \epsilon \).
In summary, several approaches have been proposed to address these issues. Strategies typically involve either embedding physical knowledge directly into the neural network architecture and formulations or improving optimization process.

\section{Experiment}
Physics-informed or physics-embedded models, enabled by advances in automatic differentiation, have become capable of approximating solutions to partial differential equations (PDEs), treating the physics as a requirement rather than a pattern learned through backpropagation. This implies that in an almost fully self-supervised model, the exact output of the network does not need to be known across the entire domain or even a part of it; yet, the network can still solve the PDE, given the initial conditions, similar to traditional PDE solvers.
We examine the performance of transport-embedded neural networks and vanilla PINN in approximating solutions to the Navier-Stokes equations on a 2D biperiodic domain. The initial conditions are based on the Taylor-Green vortex formulation.
\begin{center}
\begin{subequations}
 \label{Equations}
 \begin{align}
  \frac{\partial \bold{u}}{\partial t} + (\bold{u}.\nabla)\bold{u} &= -\frac{\nabla p}{\rho} + \frac{1}{Re}\nabla^2 \bold{u} \label{eq11} \\[1pt]
  \textbf{u}(x)_{t=0} &= \begin{bmatrix}
 \text{cos}(2\pi x)\text{sin}(2\pi y)\\
 -\text{cos}(2\pi y)\text{sin}(2\pi x)
\end{bmatrix} \label{eq12} \\[5pt]
  \nabla . \textbf{u} &= 0 
 \end{align}
\end{subequations}
\end{center}
In both models, a fully-connected neural network is designed, incorporating a prior dictionary to impose periodic boundary conditions, mapping the input $(x, y, t)$ to the output vector. For the standard PINN model, the outputs include the velocity field and pressure field ($u_x, u_y, p$). In contrast, for the transport-embedded network, the outputs consist of the vorticity and velocity fields ($\omega, u_x, u_y$), as the vorticity transport equation, derived from the Navier-Stokes equation, involves different parameters (see Eq.\ref{eq:VorttransportEq3}). Automatic differentiation is employed to calculate the partial derivatives of the outputs with respect to the inputs. The residual terms in the loss function are given by:
\begin{center}
\begin{subequations}
 \label{residuals}
 \begin{align}
    &\mathcal{L}_{\text{PDE}} := \left\| \frac{\partial \bold{u}}{\partial t} + (\bold{u}.\nabla)\bold{u} + \frac{\nabla p}{\rho} - \frac{1}{Re}\nabla^2 \bold{u}\right\|_{\bold{\Omega}}, \hspace{0.3cm} \mathcal{L}_{\text{IC,Vanilla PINN}} := \left \| \bold{u} - \bold{u(x)_{t=0}} \right \|, \hspace{0.3cm} \mathcal{L}_{\text{Incmp}} := \left\|  \text{div}(\bold{u}) \right \|_{\bold{\Omega}} \\[4pt]
  &\hspace{-0.0cm}\mathcal{L}_{\text{Curl}} := \left\| \omega - \nabla \times \bold{u} \right\|_{\bold{\Omega}}, \hspace{0.3cm} \mathcal{L}_{\text{IC,TENN}} := \left \| \bold{u} - \bold{u(x)_{t=0}} \right \|_{\bold{\Omega}} + \left \| \bold{\omega} - \bold{\omega(x)_{t=0}} \right \|_{\bold{\Omega}}, \hspace{0.3cm} \mathcal{L}_{\text{Incmp}} := \left\|  \text{div}(\bold{u}) \right \|_{\bold{\Omega}} \\[10pt]
  &\hspace{5cm}\mathcal{L}_{\text{Total}} := \boldmath{\alpha}\cdot [\mathcal{L}_{\text{PDE}}, \mathcal{L}_{\text{Curl}}, \mathcal{L}_{\text{Incmp}}, \mathcal{L}_{\text{IC,Vanilla PINN}}, \mathcal{L}_{\text{IC,TENN}}] 
 \end{align}
\end{subequations}
\end{center}
Eq.\ref{residuals}a corresponds to the residual terms for vanilla PINN when solving the Taylor-Green vortex, while Eq.\ref{residuals}b pertains to the residual terms for the transport-embedded neural network (TENN) for the same problem. Notably, the divergence-free term is the only common term between them. Eq.\ref{residuals}c defines the final loss function as a weighted sum of each residual, where $\alpha$ represents the vector of hyperparameters. When training vanilla PINN, the residual terms related to TENN have hyperparameters set to zero, and vice versa when training TENN, the vanilla PINN residual terms have zero as their hyperparameters. Also different activation functions are utilized, such as '$\text{sin}$', '$\text{tanh}$', and '$\text{softplus}$'. 
We show that our model is capable of learning a spectrum of diffusion coefficients, unlike the vanilla PINN, which struggles to avoid static solutions. In all experiments, we used the ADAM (Adaptive Moment Estimation) optimizer, differing from many previous studies that relied on the L-BFGS optimizer, a second-order method. Although L-BFGS can often provide more precise results, it operates in batch mode, which prevents stochastic training. This limitation makes learning process more prone to getting trapped in local minima, particularly in problems where the likelihood of encountering such minima is high.
\begin{figure}[!ht]%
    \centering
    {{\includegraphics[width=15.5cm]{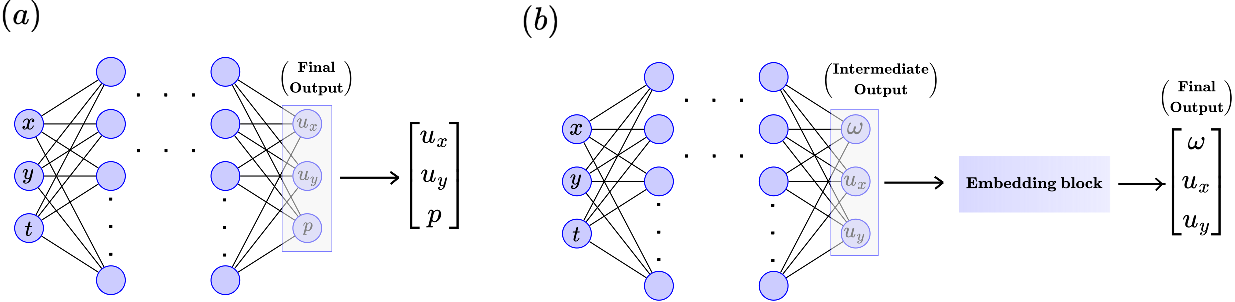}}}%
    \caption{Two neural network, in both networks the second layer is prior dictionary of periodic boundary condition shown in Fig.\ref{PD} and hidden intermediate layers(shown with dotted lines) are perceptrons with activation functions of choice, (a) vanilla PINN and (b) TENN}%
    \label{Networks}%
\end{figure}
\begin{figure}[!ht]%
    \centering
    {{\includegraphics[width=15.5cm]{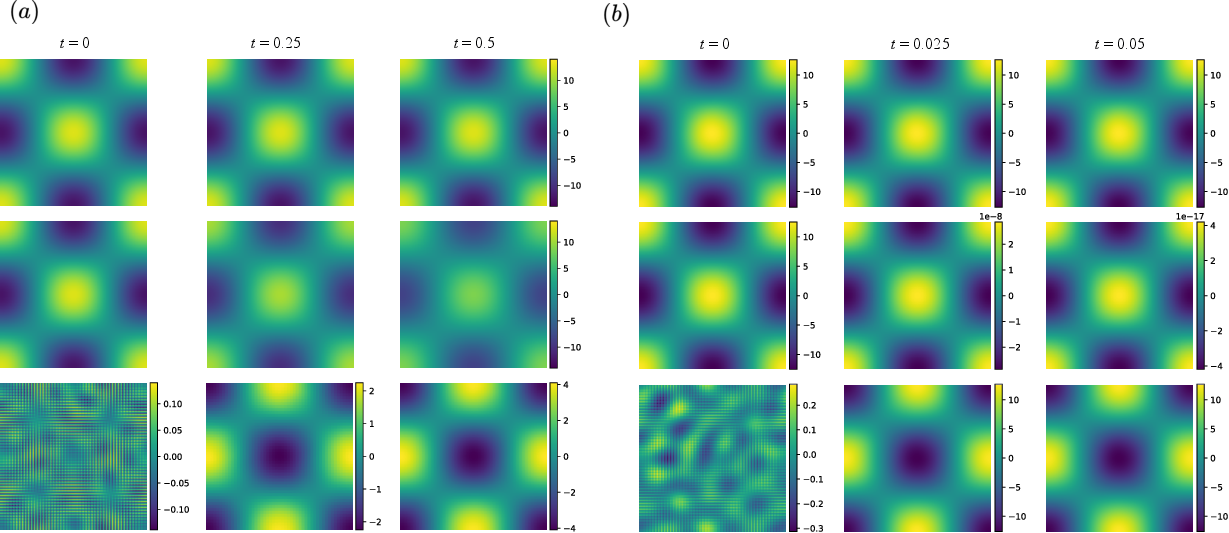}}}%
    \caption{The evolution of the vorticity field for the vanilla PINN is illustrated over time for two different Reynolds numbers: (a) \( \text{Re} = 0.1 \) and (b) \( \text{Re} = 100 \).  The top row displays the vanilla PINN predictions, the middle row shows the ground truth, and the bottom row illustrates the error between the predicted and actual results.}%
    \label{pinn-results}%
\end{figure}
In the case of the vanilla PINN, regardless of the Reynolds number, the network fails to avoid static solutions, and the maximum error consistently occurs at half the domain period, where the vorticity magnitude is minimal.
\begin{figure}[!ht]%
    \centering
    {{\includegraphics[width=15.5cm]{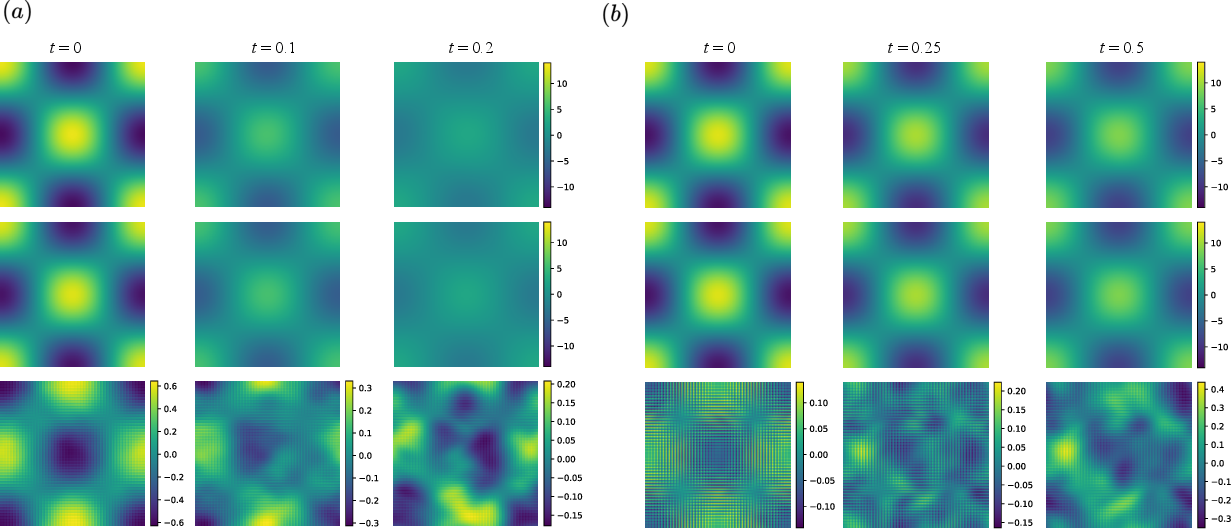}}}%
    \caption{The evolution of the vorticity field over time for the Transport-Embedded Neural Network (TENN) is shown for two Reynolds numbers: (a) \( \text{Re} = 100 \) and (b) \( \text{Re} = 10 \). The top row displays the TENN predictions, the middle row shows the ground truth, and the bottom row illustrates the error between the predicted and actual results.}%
    \label{HM-Re-100}%
\end{figure}
As shown in Fig.\ref{HM-Re-100}, for relatively high Reynolds numbers, TENN effectively captures the vortex decay dynamics, with the relative error being approximately 4\%. It is also noteworthy that as the Reynolds number increases, the error between the ground truth and TENN predictions grows over time. In both cases presented in Fig.\ref{HM-Re-100}, the maximum error occurs at the half-period of the domain, indicating that the network struggles to train when the vortex magnitude is low. As time progresses, no distinct pattern emerges in the error heatmap.

\begin{figure}[!ht]%
    \centering
    {{\includegraphics[width=15.5cm]{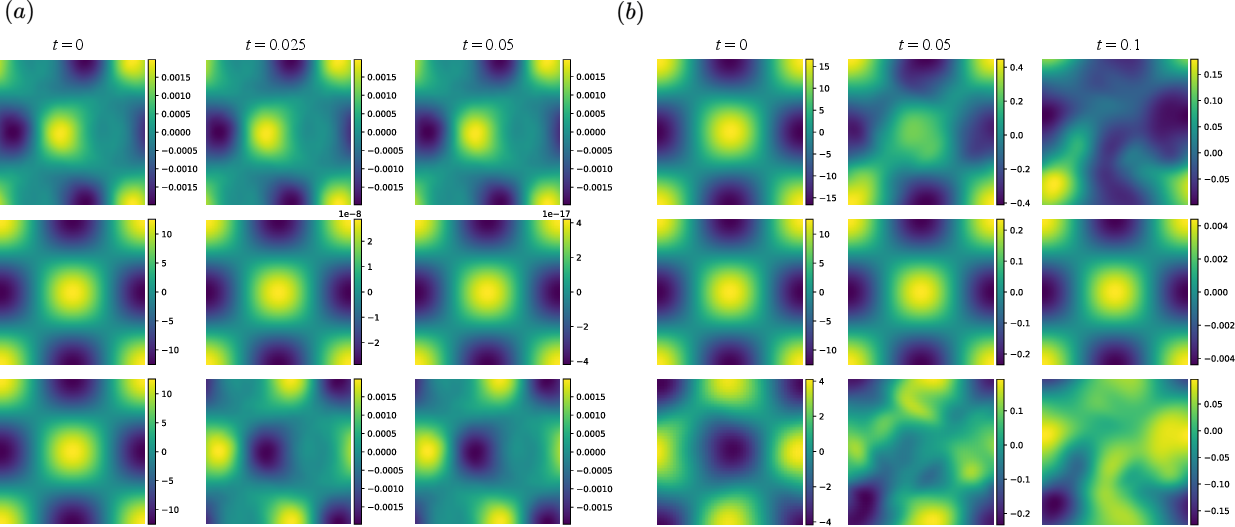}}}%
    \caption{The evolution of the vorticity field for the Transport-Enhanced Neural Network (TENN) is shown over time for two different Reynolds numbers: (a) \( \text{Re} = 0.1 \) and (b) \( \text{Re} = 1 \). The top row displays the TENN predictions, the middle row shows the ground truth, and the bottom row illustrates the error between the predicted and actual results.}%
    \label{HM-Re-1}%
\end{figure}

\begin{figure}[!ht]%
    \centering
    {{\includegraphics[width=14cm]{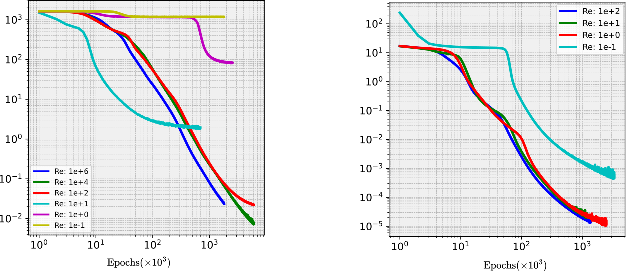}}}%
    \caption{The total loss of the neural network over epochs is depicted in the figure, with the plot on the left corresponding to the total training loss of the Transport-Embedded Neural Network (TENN), while the plot on the right represents the total loss of the vanilla Physics-Informed Neural Network (PINN).}%
    \label{total-loss}%
\end{figure} 
In the case of relatively low Reynolds numbers, as shown in Fig.\ref{HM-Re-1}, TENN successfully captures the temporal evolution of the problem. However, due to the dominance of high diffusion terms (i.e., \( \frac{\partial^2 \mathbf{u}}{\partial x_i^2} \)), the network struggles to capture the underlying physics without additional data input, highlighting its limitations as a fully self-supervised method. Additionally, as the Reynolds number increases, the error grows over time. This challenge arises from the need to compute second-order derivatives, which requires double backpropagation in the network, making it more prone to gradient vanishing. This is a fundamental issue when training networks that rely on higher-order derivatives.

The total loss functions for both the vanilla PINN and the proposed Transport-Embedded Neural Network (TENN) are presented in Fig.\ref{total-loss}. Both models encounter difficulties when training under high diffusion regimes, as discussed earlier. While the vanilla PINN achieves a low total loss, it suffers from stagnation and overfitting to the initial conditions, as depicted in Fig.\ref{pinn-results}.

\textbf{Future work and limitations}\hspace{0.2cm} This advancement opens up opportunities for solving multiphysics and multiconstrained problems, particularly in cases involving the transport of species or energy, which are traditionally challenging for machine learning approaches\cite{chen2022application}\cite{krishnapriyan2021characterizing}. By embedding physical laws directly into the network, this method allows for the resolution of complex multiphysics transport problems that previously struggled with convergence or required stringent conditions for successful computation.
\section{Conclusion}
The proposed TENN architecture represents a step forward in neural network-based CFD modeling by embedding physical transport laws directly into the neural network structure. This approach enhances the model's ability to handle convection-driven phenomena and results in smoother and more convex loss functions compared to traditional PINN methods.
However, the study also highlights the limitations of the TENN model, particularly in scenarios dominated by high diffusion effects, such as those encountered at low Reynolds numbers. In these cases, convergence was weak, and the model struggled to accurately capture the diffusion dynamics without additional data input.
In summary, the paper successfully introduces a novel neural network architecture that improves upon standard PINNs for CFD applications. Yet, the results suggest that further improvements are necessary, particularly for high-diffusion scenarios, to ensure robust performance across a wide range of flow regimes. 

\bibliographystyle{apalike}
\bibliography{references}


\end{document}